\documentclass[prl,aps,twocolumn,showpacs]{revtex4}

\usepackage{amsfonts}

\newcommand{\be}{\begin{equation}}
\newcommand{\ee}{\end{equation}}
\newcommand{\bea}{\begin{eqnarray}}
\newcommand{\beas}{\begin{eqnarray*}}
\newcommand{\eea}{\end{eqnarray}}
\newcommand{\eeas}{\end{eqnarray*}}
\newcommand{\ba}{\begin{array}}
\newcommand{\ea}{\end{array}}
\def\ls{\mathrel{\lower4pt\vbox{\lineskip=0pt\baselineskip=0pt
           \hbox{$<$}\hbox{$\sim$}}}}
\def\gs{\mathrel{\lower4pt\vbox{\lineskip=0pt\baselineskip=0pt
           \hbox{$>$}\hbox{$\sim$}}}}

\begin{document}

\title{Unification of Cosmological Scalar Fields}

\author{
Abdel P\'erez-Lorenzana\footnote{email: aplorenz@fis.cinvestav.mx},
Merced Montesinos\footnote{email: merced@fis.cinvestav.mx}
and 
Tonatiuh Matos\footnote{email: tmatos@fis.cinvestav.mx}
}
\affiliation{Departamento de F\'{\i}sica, Centro
de Investigaci\'on y de Estudios Avanzados del I.P.N.\\
Apdo. Post. 14-740, 07000, M\'exico, D.F., M\'exico}

\date{July 2007}

\begin{abstract}
We present a model where early inflation and 
late accelerating expansion of the Universe are  driven by 
the real and imaginary parts of a single complex scalar field, which we
identified as inflaton and phantom field, respectively. 
This inflaton-phantom unification is protected by an internal 
$SO(1,1)$ symmetry, with the two cosmological scalars appearing as the
degrees of freedom of a sole fundamental representation.
The unification symmetry allows to build successful potentials.
We observe that our theory  provides a matter-phantom duality, 
which  transforms  scalar matter cosmological solutions 
into phantom solutions and vice versa. We also suggest that 
a complete unification of all  scalar fields of cosmological
interest  is yet possible under a similar footing. 
\end{abstract}

\maketitle                                                               
%%%%%%%%%%%%%%%%%%%%%%%%%%%%%%%%%%%%%%%%%%%%%%%%%%%%%%%%%%%%%%%%%%%

As cosmology entered the  precision stage with the accurate
measurements of the microwave cosmic background spectrum  by
the Cosmic Background Explorer
(COBE)~\cite{cobe} and most recently  by 
the Wilkinson Microwave Anisotropy probe 
(WMAP)~\cite{wmap}, and with  the
observations of distant type Ia supernovae~\cite{pelmutter,riess,snls}
and galaxy cluster measurements~\cite{bahcall}, 
the necessity for dark components
of the Universe seems to be unavoidable. Current data indicates that common
(standard) matter barely constitutes about 3\% of the critical density of the
Universe, whereas about 27\% corresponds to dark matter (DM). The
remaining 70\% is a form of dark energy (DE), which is responsible for the 
current accelerating expansion of the Universe. 

Understanding the origin of dark components 
of the Universe has been one of the leading motivations
for a large number of theoretical works in last years. 
Apart from the cosmological constant, 
one of the favorite candidates for DE 
are scalar fields, for which acceleration is easy to achieve 
by choosing an appropriate potential energy and tuning model parameters, 
as in quintessence  and tachyonic scalar 
models~(for examples see~\cite{quintessence,tachyon}).
However, for all such models  
the equation of state $p= \omega \rho$ leads to $\omega>-1$. 
In contrast,  observations, including recent results of SNLS~\cite{snls},
not only constrain $\omega$ to be close to $\omega = -1$, 
but also seem to allow and even favor the parameter region where $\omega <-1$ 
(see also~\cite{omega}), indicating an apparent 
exotic source for DE, which violates the weak energy condition 
$\rho>0$, $\rho+p>0$~\cite{caldwell}. 
A simple way to realize this scenario is to introduce a scalar field,
$\varphi$, called phantom, 
for which the kinetic term comes with the `wrong' sign, i.e. 
${\cal L} = -\frac{1}{2} \partial^\mu\varphi\partial_\mu\varphi -
V(\varphi)$, which gives the pressure 
$p= - \frac{1}{2}\dot\varphi^2 - V$ and
the energy density  $\rho =- \frac{1}{2}\dot\varphi^2 + V$, 
leading to  $\rho+p= - \dot\varphi^2<0$. 

An increasing  number of studies has been
advocated to analyze phantom cosmology
(for references see for example~\cite{phantomrefs,phantomrefs2,lazkoz}).
Few ideas exist, however, about the possible
origin of the phantom, 
many of them as exotic as the nature of the field itself. 
They range from string motivated models~\cite{strings}
to higher order theories of gravity and supergravity~\cite{sugra}
and nonminimally coupled scalar field theories~\cite{nonminimal}.
 
The aim of this paper is  to show, first, that  despite the  
sign of its kinetic term, the phantom can actually be understood  as 
the imaginary part of a complex scalar field, $\phi$. Free $\phi$  
satisfies  standard equations of motion and thus our theory would not 
have to relay on more exotic physics.
Second, that the extra  degree of freedom which 
appears as the real part of $\phi$ 
has a standard kinetic term, and so it
could in principle play the role of any other cosmological scalar field. 
The conclusion is striking.  
One can unify in a simple way the phantom 
with some other field of cosmological use,
and there is a natural
candidate for the latter: the inflaton, which 
drove another stage of cosmological acceleration 
at the early Universe. 
Thus, in this scenario, DE would be
just a remnant of the very early stages of cosmological evolution, with the
phantom as the other face of a  more fundamental field  
(see Ref.~\cite{rosenfeld,ish} for similar ideas with quintessence). 
One can also speculate 
about  phantom and DM unification on this same theoretical ground,
but we will leave such speculations for a later discussion~\cite{arbey}.

In many areas of  physics, however, true unification is usually a very profound
concept that implies the existence of protecting symmetries that
interrelate the dynamical degrees of freedom of the theory. 
Such is the case, for instance, of electrodynamics where 
gauge invariance arises, the same  later became 
the fundamental link for
the electroweak unification in the standard model of particle physics.  
Remarkably, the theory we are 
about to develop  also fulfills this concept. 
It posses an $SO(1,1)$  symmetry 
with the inflaton and phantom belonging to a fundamental two-dimensional
representation. Once promoted to a  fundamental level in the theory, 
the symmetry  allows us to build 
successful potentials to account for inflation and late
time acceleration with the so given degrees of freedom. 
The symmetry also implies a duality  among 
matter and phantom cosmological solutions. 
Furthermore, the theory has an immediate extension
to include other scalars of cosmological interest.
Those are the central points of the present paper.

To elaborate our theory, let us start by considering 
a single complex scalar field $\phi = \frac{1}{\sqrt{2}}(\varphi_1+i\varphi_2)$.
As fundamental symmetries can usually be read out of the kinetic terms,
we will for the moment switch off any possible potential, and, contrary to
standard lore, 
we write a complex kinetic term for $\phi$, such that the 
real Lagrangian is given in the noncanonical form
 \be 
 {\cal L} = \frac{1}{2} 
 \left[\partial_\mu\phi \partial^\mu\phi + 
       \partial_\mu\phi^*\partial^\mu\phi^*\right]~;
 \label{complex}
 \ee
where the contraction of space-time indices 
with the background metric is to be understood. 
As it should be clear, despite its unusual form, 
the Lagrangian describes the correct equations of motion
for the free fields $\phi$ and $\phi^*$. 
Nevertheless, when written in terms of 
real and imaginary parts of $\phi$, 
the above Lagrangian acquires the remarkable
expression
 \be
 {\cal L} = \frac{1}{2} 
 \left[ \partial_\mu\varphi_1 \partial^\mu\varphi_1 - 
       \partial_\mu\varphi_2\partial^\mu\varphi_2\right]~;
 \label{real}
 \ee
where not only the two real component fields are decoupled, but also the
imaginary part has the opposite sign in the kinetic term with
respect to $\varphi_1$. Thus, $\varphi_2$ can indeed be identified as 
a phantom field, while the standard scalar field $\varphi_1$ still 
allows for a variety of interpretations. 
We would like to identify $\varphi_1$  
as the inflaton, thus achieving a
first step towards phantom-inflaton unification.

It is fair to mention that there already exist some models in the literature
where phantom and an ordinary scalar field are treated together on a
cosmological frame. By construction, it is clear that all such models are
described in principle by the same kinetic Lagrangian term given in Eq. (2).
Such are the cases of the so-called quintom models~\cite{quintom}, and the
hessence models~\cite{hessence}. Nevertheless, both models were constructed 
{\it ad hoc} 
to address a single problem, the nature of DE as produced by a complex
system with a phantom component. Clearly, having the same action term is just an
accidental overlapping.  The actual physics a model can account for depends not
only on the basic ingredients, but also on the extend to which they can be
identified as specific degrees of freedom of a real system.  In this sense, our
model is different to any previously presented model. We intent to guide the
construction of our cosmological model by the fundamental principles of
simplicity and the use of symmetries. And so, as we will show, it is the
identification of $\varphi_1$ with the inflaton and the identification of the
right symmetry for the model that shall provide us the simplest, meaningful, and
minimal realization of our cosmological unification idea by connecting both
stages of accelerating expansion at a very fundamental level.

Next,  we  look for the symmetries of
above kinetic terms. First, notice that the Lagrangian 
in Eq.~(\ref{complex}) can be put  
in the same functional form of Eq.~(\ref{real}), by the simple
reparametrization $\phi^*\rightarrow i\phi^*$, which resembles 
Wick transformation that connects  Euclidean to Minkowski 1+1 
dimensional spaces,  where one changes $t\rightarrow it$ 
on the Euclidean metric $ds_E^2 = dx^2 + dt^2$ to get 
$ds^2 = dx^2 - dt^2$. 
As a matter of fact, the analogy has a deeper meaning which 
becomes transparent if we define the 
two-dimensional vectorlike array 
 \be 
 \Phi_E = \left(\ba{l} \phi\\ \phi^*\ea\right)~,
 \label{phie}
 \ee
in terms of which Eq.~(\ref{complex}) 
is simply written as the
Euclidean metric contraction 
\be 
{\cal L} = \frac{1}{2} \partial_\mu\Phi_E^T\cdot \partial^\mu\Phi_E 
 \label{le}
\ee
on field space (aside from derivatives), where $T$ stands for the transpose.
Similarly, the Lagrangian in Eq.~(\ref{real}) can be expressed as 
\be 
{\cal L} = \frac{1}{2} \partial_\mu\Phi^T\cdot \sigma_3\cdot \partial^\mu\Phi
= \frac{1}{2} \,\eta_{ij}\,\partial_\mu \Phi^i\,\partial^\mu \Phi^j~, 
\label{lm}
\ee
where we have now written the real field components as
 \be 
 \Phi = \left(\ba{l} \varphi_1\\ \varphi_2\ea\right)~,
 \label{phi}
 \ee
and  we have recognized the 1+1 dimensional 
Minkowski metric as $\eta=\sigma_3={\rm diag}(1,-1)$. 

The conclusion is straightforward. The simple 
Lagrangian theory we have just presented for
inflaton and phantom unification has an internal guarding symmetry. 
That corresponds to the symmetry transformations on field space
that preserve the 1+1 Euclidean (or Minkowski) metric. 
For the Euclidean contraction on Eq.~(\ref{le}) we have 
the symmetry group $O(2,{\mathbb{C}})$, consisting of all two by two 
orthogonal complex  matrices. 
$O(2,\mathbb{C})$ is a  dimension one, nonconnected, complex Lie group, 
which is isomorphic to the 
indefinite orthogonal group $O(1,1,\mathbb{C})$,
whose transformations preserve the 
Minkowski metric contraction in Eq.~(\ref{lm}). 
The identity component of $O(2,\mathbb{C})$, i.e. the subgroup of
symmetry transformations connected to the identity, is 
$SO(2,\mathbb{C})$ which is isomorphic to 
$SO(1,1,\mathbb{C})\subset O(1,1,\mathbb{C})$.
The  algebra of the former 
has a single generator over $\mathbb{C}$, that one identifies
as $i\sigma_2$, whereas the generator of $SO(1,1,\mathbb{C})$ is $\sigma_1$. 
$SO(1,1,\mathbb{C})$ contains two Lorentz-like subgroups $SO(1,1)$, which 
by using the exponential mapping are expressed as 
the complex rotations $g_\alpha = \exp(i\,\alpha\, \sigma_1)$ 
and the standard (real) Lorentz boosts 
$h_\alpha = \exp(\alpha\, \sigma_1)$; for $\alpha$ real.
$\Phi$ corresponds 
to the fundamental (two-dimensional complex)
representation. It is easy to see that  the Wick transformation 
$\phi^*\rightarrow i\phi^*$ on $\Phi_E$ gives 
\be
\Phi_E\rightarrow 
\left(\ba{c} \phi\\ i\phi^*\ea\right)~ = 
e^{i(\pi/4)\sigma_1}\left(\ba{l} \varphi_1\\ \varphi_2\ea\right)~,
\ee
which shows the equivalence of both prescriptions. 

Given any two arbitrary doublet
representations,  $\Phi$ and $\Psi$, 
there is only one invariant bilinear form under  $O(1,1,\mathbb{C})$ 
transformations, $\Phi^T\sigma_3\Psi=\eta^{ij}\Phi_i\Psi_j$. 
The reduced group $SO(1,1,\mathbb{C})$, however,
has the extra invariant  
$\Phi^T\,i\sigma_2\Psi=\epsilon^{ij}\Phi_i\Psi_j$, where 
$\epsilon^{ij}$ is the usual skewsymmetric tensor with $\epsilon^{12}=1$. 
Two extra invariants  exist if we restrict to any of the 
$SO(1,1)$ subgroups. $g_\alpha$  complex
rotations have also the invariants  
$\Phi^\dagger\sigma_1\Psi = \Phi_1^*\Psi_2 +\Phi_2^*\Psi_1$; 
and  $\Phi^\dagger\Psi = \delta^{ij}\Phi_i^*\Psi_j$.
On the other hand, 
$\Phi^\dagger\,i\sigma_2\Psi $; and $\Phi^\dagger\sigma_3\Psi$ are 
invariant under the real boost transformations.
However, notice that with the real component  field representation given in 
Eq.~(\ref{phi}), 
the last invariants are not independent
from those of $SO(1,1,\mathbb{C})$.

We will choose the $SO(1,1)$
subgroup given by $g_\alpha$
transformations as the fundamental symmetry of our theory. 
This has the clear advantage that one can directly work with 
the two real components 
by using $\Phi$ as given in Eq.~(\ref{phi}), 
thus keeping our initial identification
for the phantom and inflaton candidate fields. 
Therefore, our theory explicitely breaks the $O(1,1,\mathbb{C})$ isometry group
down to an $SO(1,1)$ residual symmetry through the potential terms. 
Notice that this is unlike hessence models in Ref.~\cite{hessence} 
where the other $SO(1,1)$ subgroup was used.

Clearly, since $\Phi^T\,i\sigma_2\Phi=0$,
we may only use the other three  associated invariants 
to build the potential of our
unified theory, which we may rewrite as
 \bea 
 \Phi^T\sigma_3\Phi &=& \varphi_1^2 - \varphi_2^2~; \\
 \Phi^\dagger\Phi &=& \varphi_1^2 + \varphi_2^2~; \\
 \Phi^\dagger\sigma_1\Phi &=& 2\varphi_1\varphi_2 ~. 
\eea 
It is worth noticing that the first two expressions above 
allow us to write quite general
and independent potentials for phantom and inflaton fields 
as far as they depend quadratically on such fields: 
$U(\varphi_1^2) = U(\Phi^T\sigma_3\Phi + \Phi^\dagger\Phi)$ and 
$V(\varphi_2^2) = V(\Phi^\dagger\Phi-\Phi^T\sigma_3\Phi)$, 
respectively.
In this scenario, phantom and inflaton have consistently decoupled
dynamics, provided the coupling term 
$\Phi^\dagger\sigma_1\Phi$ is not allowed. Remarkably, we can again make use of
symmetry arguments to insure this feature by requiring the theory to be
invariant under the parity transformation generated by the metric
$\eta=\sigma_3$, under which  $\Phi^\dagger\sigma_1\Phi$ 
is a pseudoscalar. Other bilinears remain invariant.
Thus,  a natural and simple choice is to take
a chaotic potential for the inflaton,
$U(\varphi_1) = \frac{1}{2}\mu^2\varphi_1^2$. 
For the phantom field, on the other hand, there is not a preferred
potential yet, but  we can include most working examples in the
literature (see~\cite{phantomrefs,phantomrefs2} for references). For instance, 
one can consider the toy phantom potential with a  bell profile 
 \be
 V_0~\exp\left(-\frac{\alpha}{M_P^2}\varphi_2^2\right)~.
\label{expv}
 \ee

The dynamics with such potentials 
follows the general features of scalar
phantom  cosmology. 
Particularly, phantom obeys the equation of motion
 \be 
\ddot\varphi_2 +3H\dot\varphi_2 =\frac{\partial V}{\partial\varphi_2}~,
\label{eqmotion}
 \ee
which indicates that the expansion of the Universe acts as a damping force, as
usual, but the phantom moves towards local maxima, as if the potential were
inverted, due to the  sign on the RHS of Eq.~(\ref{eqmotion}).
Initial 
conditions are expected to be  fixed at same time for both scalars. 
Early inflation would
proceed as usual, followed by  reheating, big-bang nucleosynthesis, structure
formation, and a late time matter-dominated era. 
The phantom should survive all those epochs
without affecting them. This can be arranged by fine tuning the 
potential parameters such that the phantom would remain
frozen at some large value away from zero, until matter density 
$\rho_m$ catches up with it at late time. 
Indeed, if initial phantom conditions can be arranged for almost zero initial
kinetic energy, $T=-\dot\varphi_2^2/2$, and 
a  very flat  potential (small $\alpha$ regime),
the phantom starts with an equation of state with
$\omega = -1$, with $\rho_m\gg\rho$ during matter domination. 
As the Universe expands, matter density
scales as $\rho_m\propto a^{-3}$, with $a$ the metric scale function, 
whereas phantom density goes as $\rho\propto exp[-6\int\!(1-\theta(a))da/a]$,
for $\theta(a) = (1+T/2V)^{-1}$. Flatness of the potential would imply that
$\theta\approx 1$, well up to the coincidence era, where
 $\Omega_{\rm matter}$ equals $\Omega_{\rm phantom}$.
Thereafter, phantom gets released, gaining kinetic energy and 
moving  towards the top of the potential producing the 
$\omega<-1$ era, where we live. 
This would be part of a period of
phantom oscillations around the maximum of the potential, 
where the phantom should finally
settle, returning to $\omega=-1$ within finite time.
In this epoch, the Hubble parameter gets contributions from both sources, 
$H^2=(\rho+\rho_m)/3M_P^2$. $\theta$
becomes larger than one, and phantom energy increases  with the
expansion whereas matter keeps diminishing.
Thus, it is natural to expect that $\Omega_m<\Omega_{\rm DE}$ as
observations indicate.
Actual values can be accommodated, including supernova data, for some models 
(see for instance Ref.~\cite{phantomrefs}).

Alternative unified scenarios where the two field components are not
completely decoupled are also possible. Those are, to our point of view 
quite more interesting. 
Consider, for instance, the simple potential
 \be
  U(\Phi) = \frac{1}{4}M^2\left( \Phi^T\sigma_3\Phi + \Phi^\dagger\Phi\right)
   - \frac{1}{2} m^2 \Phi^\dagger \sigma_1\Phi + V_0~,
 \ee
built out of the three $SO(1,1)$ invariants. 
It can be written in terms of real
component fields as 
 \be
  U = 
  \frac{1}{2} M^2\left(\varphi_1 -\frac{m^2}{M^2}\,\varphi_2\right)^2 
  - \frac{1}{2} \frac{m^4}{M^2}\,\varphi_2^2 + V_0~.
 \ee 
The potential is unbounded, but
this should not be a matter of concern due to unusual dynamics of the phantom.
Above potential has a saddle point at $\Phi =0$, 
which is a local minimum for
$\varphi_1$ but a local maximum for $\varphi_2$. 
This is the stable
configuration point where the fields should finally settle down, no matter what 
the initial conditions are.  
This point is an attractor for the system. 

For $M\gg m$, the potential is steeper along $\varphi_1$, with an almost flat
direction along $\varphi_1=\frac{m^2}{M^2}\,\varphi_2$. 
If initial conditions are such that
$\varphi_{1}$ and $\varphi_{2}$ are about $M_P$,  
with  no kinetic energy,
we would be in a situation where
the dynamics of the system is initially reduced to  
chaotic inflation, with $\varphi_1$ as the inflaton. $\varphi_1$ shall roll 
down the potential towards the
local minimum  at $\varphi_1 =\frac{m^2}{M^2}\,\varphi_2$, inflating and
then reheating the Universe, whereas 
$\varphi_2$ remains frozen due to its small effective  mass, 
$\mu =  m^2/M\ll H\approx M$.
It is clear that to produce the observed 
amount of  density perturbations in the cosmic microwave background, 
we require $\delta\rho/\rho\approx M/M_P\approx10^{-5}$, which fixes the 
scale $M$.
After the period of reheating, the system settles at the local
minimum, where the effective potential becomes
$V = V_0(1-\alpha\varphi_2^2/M_P^2)$, with 
$\alpha = \frac{1}{2}m^4M_P^2/V_0M^2$. This potential has the form
of the very first terms in the expansion of the potential in 
Eq.~(\ref{expv}), which is actually true for almost 
any bell-shaped potential. 
Thus, dynamics should follow similar paths.
The system would remain at such a point 
for  a long period until the condition $\rho<\rho_m$ is broken
when the system is again released. 
Thereafter, the configuration shall move towards zero 
with a phantom dynamics, controlled by the rolling of $\varphi_2$, which 
undergoes
damped oscillations around zero. 
Time evolution of $\omega$  would help to avoid 
future singularities caused by the instabilities 
associated to the  violation of
null energy conditions (see first references in ~\cite{phantomrefs2}).  
A detailed numerical analysis (that will be
presented elsewhere) should probe this potential and fix the 
free parameters to reproduce data. 
By comparing with the analysis in Ref.~\cite{phantomrefs}, it is not difficult
to realize that the values $V_0\approx 10^{-44}~GeV^4$ and 
$m\approx 5.5\times10^{-13}~eV$, are likely to provide a successful scenario,
where the initial condition on $\varphi_2$ is chosen to give a  positive $H^2$. 
The smallness of those parameters indicates the need for a large fine
tuning, not protected by symmetry, as in the   
cosmological constant problem.

Reheating by inflaton decay can also occurs 
in a way that does not break the protecting
symmetry of our theory. Consider 
a singlet $\psi_0$ and a doublet $\Psi^T=(\psi_1,\psi_2)$ of fermions 
with the  $SO(1,1)$ invariant couplings
\be
\alpha \bar\psi_0\left(\Phi^T\sigma_3\Psi + \Phi^\dagger\Psi\right) + 
\beta \bar\psi_0\left(\Phi^Ti\sigma_2\Psi + \Phi^\dagger\sigma_1\Psi\right)~,
\ee
which, written in terms of field components, reduce to 
$2\alpha\bar\psi_0\varphi_1\psi_1 + 2\beta\bar\psi_0\varphi_1\psi_2$. 
Thus, one gets  the  
decay channels $\varphi_1\rightarrow\psi_0\psi_1,~\psi_0\psi_2$
for the inflaton,
with no coupling among phantom and fermion fields.
We then get the reheating temperature 
$T_r\approx 6\times 10^{-3}\max\{|\alpha|,|\beta|\}M_P$.

General models based on our $SO(1,1)$ symmetry, regardless  of  
the identification of $\varphi_1$, also have an interesting property  
that it is worth commenting on.
Consider again our initial analogy with the 1+1 metric.
By performing the
exchange $t\leftrightarrow x$  we get a conformally equivalent space-time
theory, since $ds^2\rightarrow -ds^2$. In field space, this
corresponds to the duality transformation  $\Phi\rightarrow \sigma_1\Phi$, 
which exchanges $\varphi_1\leftrightarrow\varphi_2$.  
$SO(1,1,\mathbb{C})$ invariant forms are odd
under this transformation, whereas the other two $SO(1,1)$ invariants 
are even. Thus, one can use this duality to transform matter into phantom
cosmological solutions, and vice versa, just by 
rewriting all terms in an invariant way.
A similar duality was noticed in
Refs.~\cite{lazkoz}, but no connection to the underlying symmetry was made.

Last, but not least, our theory can straightforwardly be extended 
to provide the possibility of a completely unified
treatment of all cosmological scalars.  Briefly,
by considering the kinetic terms for $n$ scalar fields and a phantom,
we can write them as the $n+1$ metric form 
$\frac{1}{2} \,\eta_{ij}\,\partial_\mu \Phi^i\,\partial^\mu \Phi^j$, with the
obvious $O(n,1,\mathbb{C})$ symmetry, and $\Phi$ representation. 
A model for hybrid inflation, DM and DE unification would then have 
the suggestive
symmetry $O(3,1,\mathbb{C})$, which contains the  subgroup
$SO(3,1)$. We will address these issues in a forthcoming paper.

Summarizing, we have shown that the 
phantom field that may cause current cosmic acceleration can  
be unified with 
the inflaton field that drove early expansion of the Universe, 
as the imaginary and real parts, respectively, 
of a more fundamental complex scalar field. 
The theory that describes the unification is protected by an 
internal $SO(1,1)$  symmetry.  Therefore, one can achieve
unification in a true sense, with the inflaton-phantom system belonging 
to a fundamental representations of an internal symmetry group. 
The symmetry allows us to write 
adequate invariant potentials with enough freedom to get a successful
description of both inflation and DE from the same footing.
To our knowledge, this is the very first time that the concept of 
unification via this nontrivial symmetry 
is realized for the cosmological setup. 
 
This work was supported in part by CONACyT, M\'exico, grant numbers 
49865-F, 54576-F, 56159-F,  and I0101/131/07 C-234/07. 
APL and TM are  part of the 
Instituto Avanzado de Cosmolog\'{\i}a (IAC) collaboration.

%%%%%%%%%%%%%%%%%%%%%%%%%%%%%%%%%%%%%%%%%%%%%%%%%%%%%%%%%%%%%%%%%%%%%
\vskip10pt
%%%%%%%%%%%%%%%%%%%%%%%%%%%%%%

%%%%%%%%%%%%%%%%%%%%%%%%%%%%%%

\end{document}